\documentclass[aps,prl,twocolumn,superscriptaddress]{revtex4}
\usepackage{mathrsfs}
\usepackage{epsfig}
\usepackage{graphicx}
\usepackage{amsfonts}
\usepackage[figuresright]{rotating}
\usepackage{amssymb}
\usepackage{amsmath}
\usepackage{dcolumn}
\usepackage{bm}
\usepackage{color}

\def\be{\begin{equation}} \def\ee{\end{equation}}
\def\bea{\begin{eqnarray}} \def\eea{\end{eqnarray}}

\newcommand{\WQCASQC} {Wilczek Quantum Center and Shanghai Research Center for Quantum Sciences, School of Physics and Astronomy, Shanghai Jiao Tong University, Shanghai 200240, China}

\newcommand{\sjtu} {Key Laboratory of Artificial Structures and Quantum Control,
Shanghai Jiao Tong University, Shanghai 200240, China}

\begin{document}
%\title{Imaginary time crystal: {\clb an exotic phase of quantum matter}}
\title{Vacuum induced three-body delocalization in cavity quantum materials }

\author{Zhengxin Guo}
\affiliation{\WQCASQC}

\author{Zi Cai}
\email{zcai@sjtu.edu.cn}
\affiliation{\WQCASQC}
\affiliation{\sjtu}

\begin{abstract}  % Science journal - 125 word or less; current 119.
In this study, we demonstrate that the vacuum itself suffices to delocalize an Anderson insulator inside a cavity. By studying a disordered one-dimensional spinless fermion system coupled to a single photon mode describing the vacuum fluctuation, we find that even though the cavity mode does not qualitatively change the localization behavior for a single-fermion system, it indeed leads to delocalization via a vacuum fluctuation-induced correlated hopping mechanism for systems with at least three fermions. A mobility edge separating the low-energy localized eigenstates and the high-energy delocalized eigenstates has been revealed.  It is shown that such a one-dimensional three-fermion system in with  correlated hopping can be mapped to a single-particle system with hopping along the face diagonals in a three dimensional lattice.    The effect of the dissipation as well as a many-body generalization have also been discussed.

\end{abstract}

%\pacs{05.30.Jp, 75.10.Pq, 02.70.Ss, 03.65. Yz}

\maketitle

{\it Introduction --} A vacuum is not truly empty, but instead contains  fields and particles  blinking in and out of existence for a fleeting moment. This phenomenon, known as vacuum quantum fluctuation, is responsible for a wealth of profound phenomena ranging from Casimir effect\cite{Casimir1948} to spontaneous emission\cite{Dalibard1982}. Recently, control of quantum vacuum fluctuation in nanoscale microcavities  has been used to manipulate the properties of quantum materials inside the cavity\cite{Riek2015,Benea-Chelmus2019}, giving rise to a new field coined  ``cavity quantum material''\cite{Sentef2022}.    Compare to conventional quantum manipulation  method via ultrafast classical irradiation (also known as Floquet engineering\cite{Oka2018,Disa2021}),  employing cavity vacuum with quantized electromagnetic fluctuations takes the advantage that even in a dark cavity without phonon, the pure quantum  vacuum fluctuations suffice to significantly increase the light-matter coupling via squeezing the cavity volume, thus access an ultrastrong or deep strong coupling regime where the properties of the quantum materials can be drastically modified\cite{Diaz2019,Kockum2019,Vidal2021}. Remarkable examples  include the cavity-enhanced superconductivity\cite{Sentef2018,Schlawin2019,Chakraborty2021} and breakdown or enhancement of topological protection in quantum Hall system\cite{Appugliese2022,Enkner2024}, as well as engineering the band structures\cite{Nguyen2023,Ke2023,Jiang2024,Yang2024} and transport properties\cite{Hagenmuller2018,Rokaj2022,Eckhardt2022,Arwas2023} of quantum materials.

The discovery of Anderson localization\cite{Anderson1958}, originating from the interference effect of classical or matter waves, is a remarkable progress in solid state physics. It has been generalized to interacting quantum systems, and intensively studied in the context of many-body localization (MBL)\cite{Basko2006,Oganesyan2007,Znidaric2008,Pal2010}. Since a system is inevitably coupled to its surroundings, understanding quantum localized systems immersed in an environment is not only of fundamental interests\cite{Nandkishore2014,Johri2015,Huse2015,Luitz2017,Li2024}, but also with practical significance due to its relation with realistic experiments\cite{Schreiber2015,Luschen2017, Haiming2023}. Specific to the cavity quantum materials, the cavity photons play the role of environment, but with much fewer degrees of freedom than those of conventional heat bath. Such a small bath composed of single-mode photons uniformly coupled to the whole system, mediates the interaction between the particles, and results in effective long-range interactions or hoppings that may qualitatively change the nature of the quantum localization, and give rise to new type of delocalization mechanism.

\begin{figure}[htbp]
\centering
\includegraphics[width=0.5\textwidth]{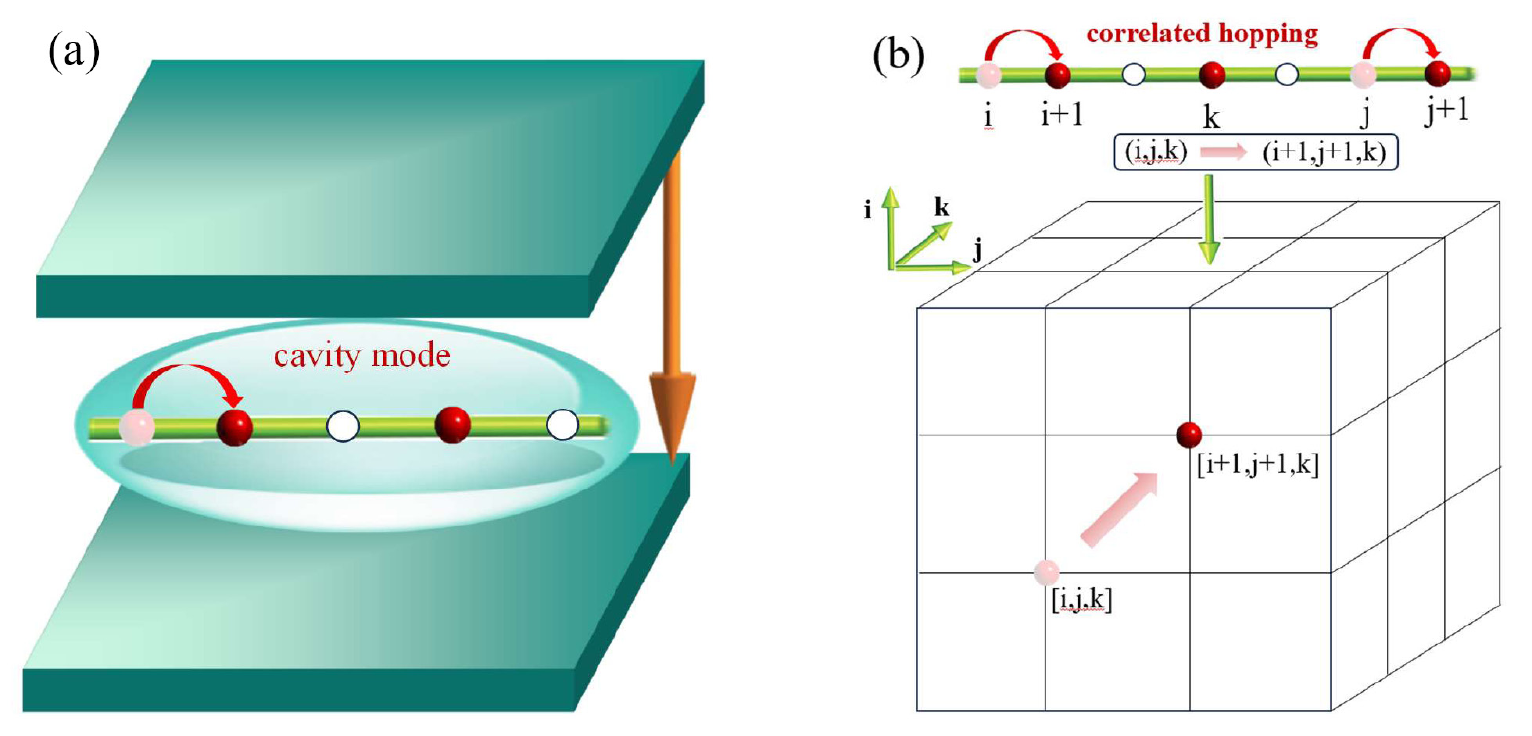}
\caption{(a) Schematic drawing of the one-dimensional fermionic system inside a cavity. (b)An example of the correlated hopping in a three-body system in a 1D lattice (upper) and its mapping to a single-particle hopping along the face diagonal of a 3D cubic lattice (lower).}
\label{fig:fig1}
\end{figure}

\begin{figure*}[htb]
\includegraphics[width=0.9\textwidth]{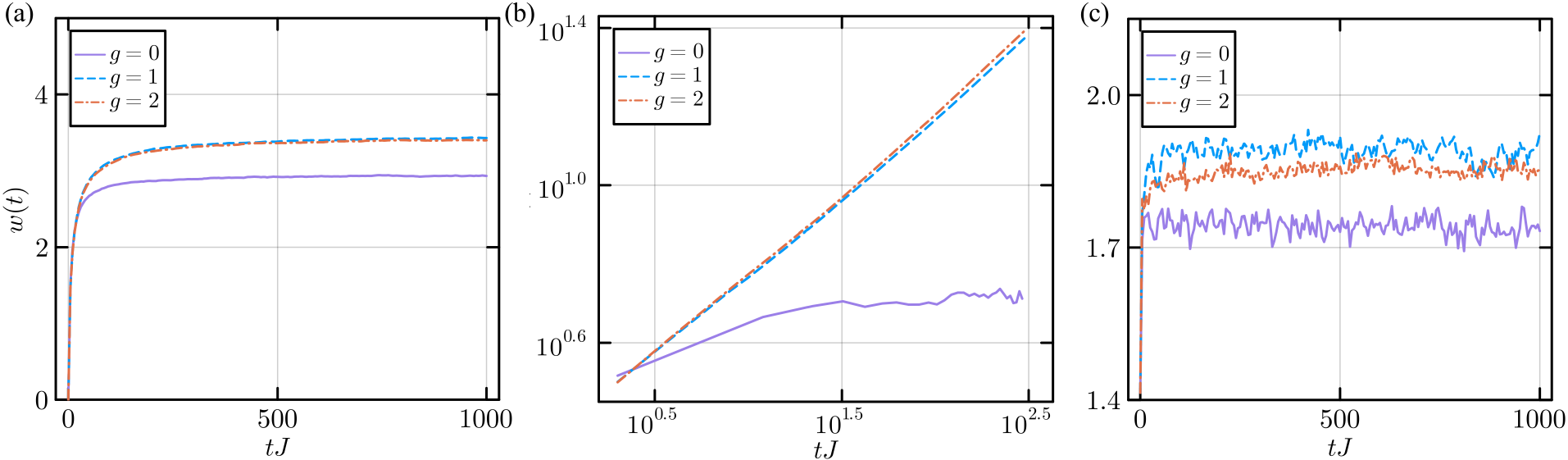}
\caption{(a) Single-particle and three-particle dynamics of the wavepacket variance $\omega(t)$ with (b) weak  and (c) strong disorder in the presence of different fermion-photon coupling strengths  $g$. System parameters for (a)-(c) are the chosen as $L=100$, $N^b_{max}=2$, $\omega_c=J$, $\mathcal{N}_{traj}=1000$. $\Delta=3J$ for (a) and (b) and $\Delta=15J$ for (c). The initial states are chosen as $|\psi\rangle^{1(3)}(t=0)=|\psi\rangle_b\otimes |\psi\rangle^{1(3)}_f$, where $|\psi\rangle_b$ is the photon vacuum state for both (a)-(b), while $|\psi\rangle^{1}_f=|0_1\cdots 1_{i_0}\cdots0_L\rangle$ for (a) and $|\psi\rangle^3_f=|0_1\cdots 1_{i_0-1}1_{i_0}1_{i_0+1}\cdots0_L\rangle$ for (b) and (c), $i_0=L/2$.}
\label{fig:fig2}
\end{figure*}

In this study, we investigate the effect of the vacuum fluctuation on the localization of a  one-dimensional (1D) non-interacting fermionic system. It is shown that even though the vacuum fluctuation does not change the nature of the single-particle localization, it could lead to  delocalization for a system with at least three fermions, via a correlated hopping mechanism. The properties of the eignestates have been studied, and a mobility edge separating the low-energy localized states and high-energy extended states has been observed in such a three-body Hamiltonian. To understand this phenomenon, we provide a mapping from such a 1D three-body system with correlated hopping to a three-dimensional (3D) single-particle system with regular hopping, which exhibits the Mott transition.    A generalization of this picture to  many-body case and the effect of dissipation due to the leakage of the photons from the cavity have also been discussed.

{\it Model and method --} The full Hamiltonian of the cavity-fermion coupled system reads:
\begin{equation}
H=\omega_c a^\dag a+\sum_i\{-J[e^{i g (a+a^\dag)}c_i^\dag c_{i+1}+h.c]+\mu_i n_i\} \label{eq:ham}
\end{equation}
where $\omega_c$ is the frequency of the single cavity mode, $a^\dag$/$a$ is the creation/annihilation operator of the photon in this cavity mode related to the quantized electromagnetic vector potential. $c^\dag_i$/$c_i$ is the fermionic creation/annihilation operator on site i of a 1D lattice with length $L$ (the lattice constant is set as unit). $J$ indicates  the amplitude of the single-particle hopping between adjacent sites and $g$ parameterizes the cavity-fermion coupling strength. $n_i=c_i^\dag c_i$ is the local density operator on site i, and $\mu_i\in [-\Delta,\Delta]$ is a random number with box distribution representing the disordered onsite energy.

We study the properties of Hamiltonian.(\ref{eq:ham}) using the exact diagonalization. For a single-mode spatially uniform vector potential, it has been proved  photon condensation is prohibited\cite{Bacciconi2023}, which allows us to truncate the Hilbert space of the photon by introducing a photon number cutoff $N_b$ (The effect of the photon number cutoff has been discussed in the Supplementary material (SM)\cite{Supplementary}.  The total number of fermion $N_f$ is conserved in Hamiltonian.(\ref{eq:ham}). Therefore, the total Hilbert space of Hamiltonian.(\ref{eq:ham}) is $\frac{N_f!(N_b+1)}{L!(L-N_f)!}$. We perform the ensemble average over $\mathcal{N}$ disorder realizations.

{\it Localization in the single-particle dynamics --} We first consider the simplest case with only one fermion in the lattice ($N_f=1$). The initial state is prepared as a product state of fermion and photon $|\psi(0)\rangle=|\psi\rangle_b\otimes |\psi\rangle_f$, where the photon is in its vacuum state $a|\psi\rangle_b=0$, and fermion is in a product state with only one fermion placed in the lattice center $|\psi\rangle_f=|0_1\cdots 1_{i_0}\cdots0_L\rangle$ with $i_0=L/2$. Starting from such an initial state, we will monitor dynamics of the fermion via the variance of its wavepacket $\omega(t)=\langle \sqrt{\sum_i (i-i_0)^2 n_i(t)}\rangle_\xi$ with $n_i(t)=\langle \psi(t)|n_i|\psi(t)\rangle$ being the onsite density of the fermion at time t, and the ensemble average $\langle\rangle_\xi$ is performed over the independent disorder realizations $\{\mu_i\}$. $\omega(t)$ with different coupling strength $g$ are plotted  in Fig.\ref{fig:fig2} (a), which shows that $\omega(t)$ quickly saturates to a finite value, a signature of Anderson localization. It shows that the vacuum fluctuations subject to a single fermion do not qualitatively change its localization picture compared to the $g=0$ case, even though it  indeed changes the localization length as well as the conductivity\cite{Hagenmuller2018}.

{\it Delocalization and Mott transition in a 1D three-body problem --} The situation is qualitatively different if two other fermions are added. To verify this, we keep all the conditions  intact ({\it e.g.} system parameters and the initial state of boson), but only change  the initial state of fermion as three fermions placed in the lattice center $|\psi\rangle_f=|0_1\cdots 1_{i_0-1} 1_{i_0}1_{i_0+1}\cdots0_L\rangle$. We still focus on the dynamics of the variance of three-fermion packet $\omega(t)=\langle \sqrt{\sum_i [i-i_0]^2 n_i(t)}\rangle_\xi$. Fig.\ref{fig:fig2} (b) indicates that for $g=0$, $\omega(t)$ approaches a saturated value, a signature of Anderson localization irrespective of the disorder strength.  However, coupling to the photon vacuum  ($g>0$) qualitatively changes this picture for a weak disorder, and leads to a subdiffusion: $\omega(t)\sim t^{\alpha}$ with an disorder-dependent exponent $\alpha<0.5$. As the disorder strength further increases, the three-body wavepacket localizes again, as shown in Fig.\ref{fig:fig2} (c), indicating a delocalized-to-localized transition for such a 1D system. It is shown that the universality class of this phase transition in such a 1D model agrees with that of the Mott transition in a 3D Anderson localization\cite{Supplementary}.

 \begin{figure}[htbp]
\centering
\includegraphics[width=0.49\textwidth]{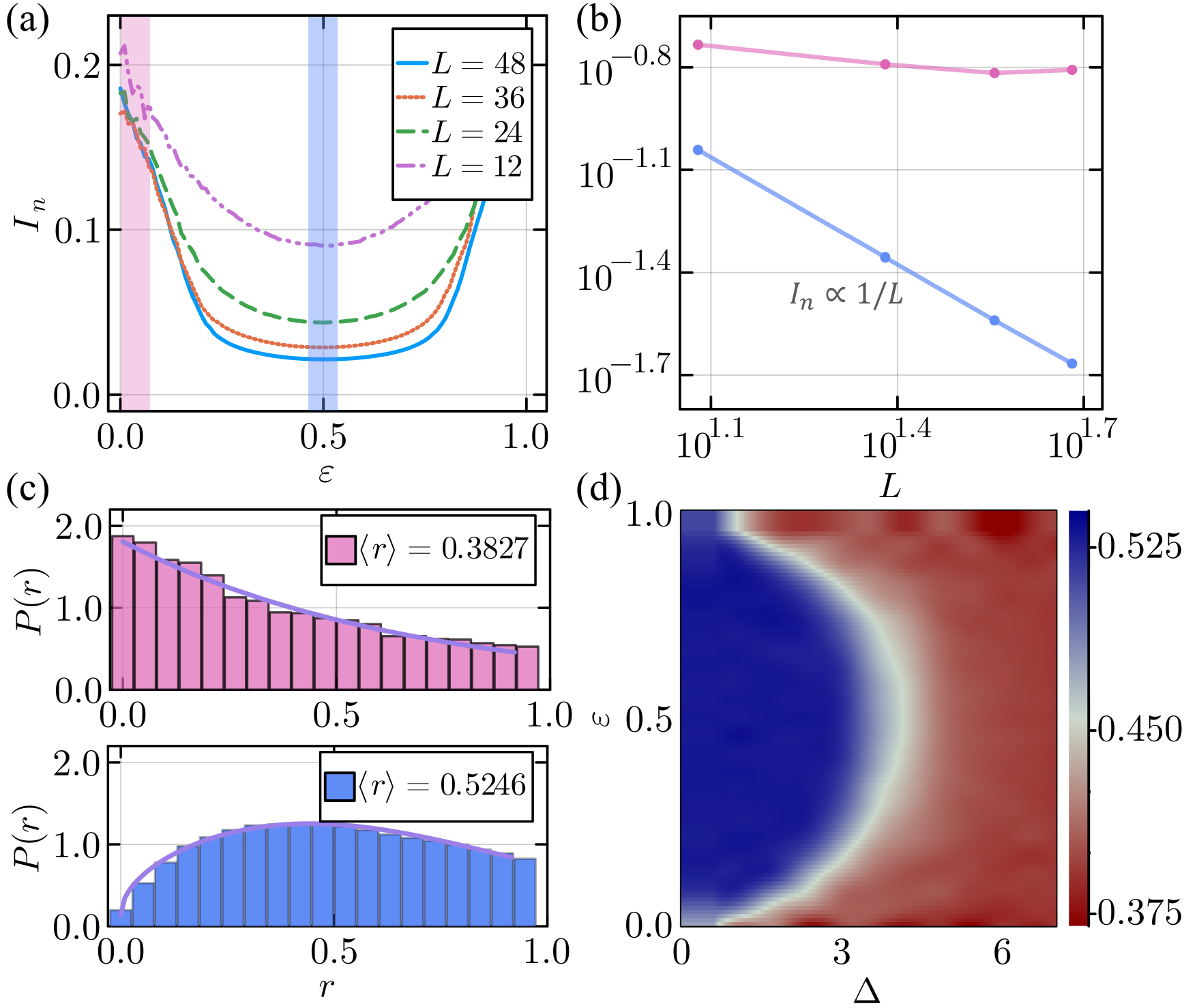}
\caption{(a) $I_n$ as defined in Eq.(\ref{eq:In}) for all eigenstates $|n\rangle$ of a three-body Hamiltonian as a function of its normalized energy $\epsilon$ for different system sizes. The red and blue area indicate the eigenstate regimes whose normalized energies satisfy  $\epsilon\in [0.0.04]$ (spectrum edge) and $\epsilon\in [0.48,0.52]$ (spectrum center)  respectively, representing the typical regimes with localized and delocalized eigenstates. (b) The average $I_n$ over the eigenstates in blue and red regimes in (a) as functions of $L$, which exhibit signatures of delocalization $I_n\sim 1/L$ (blue dot) and localization $I_n\sim \mathcal{O}(1)$ (red dot) respectively. (c)The statistics of the ratio of adjacent gaps in the red(upper panel) and blue(lower panel) regimes of the spectrum, which agree with the Poisson and GOE distributions (the solid line) respectively. (d)The average value of the ratio of adjacent gaps as a function of $\epsilon$ and disorder strength $\Delta$. The parameters are chosen as $g=1$, $N_f=3$, $N_b=2$, $\omega_c=J$  for (a)-(d), $\Delta=3J$ for (a)-(c) and $L=36$ for (c) and (d). $\mathcal{N}_{traj}$ ranges from $10^3$ to $10^4$ depending on $L$.}
\label{fig:fig3}
\end{figure}

Such a delocalization can be understood as a consequence of a long-range correlated hopping induced by integrating the photon degrees of freedom. To see this, we consider the weak coupling limit $g\ll 1$, where $e^{ig(a+a^\dag)}\simeq 1+ig(a+a^\dag)$, and the partition function of the whole system reads: $Z\simeq Tr e^{-\beta(H_s+ H_b+H_{sb})}$, where
\begin{equation}
H_s=\sum_i[-J(c_i^\dag c_{i+1}+h.c)+\mu_i n_i], \label{eq:sysHam}
\end{equation}
is the system Hamiltonian. $H_b=\omega_c b^\dag b$, $H_{sb}=gJ (a^\dag+a) \sum_i J_{i,i+1}$ where $J_{i,i+1}=i(c_i^\dag c_{i+1}-c^{\dag}_{i+1}c_i)$ is the current operator in the bond $[i,i+1]$. In the path integral formalism\cite{Feynman1976}, one can integrate out the photon degrees of freedom, and derive an effective action as:
\begin{equation}
Z=Tr e^{-\beta [H_s+H_b+H_{sb}]}=Z_b  \int \mathcal{D}c^\dag_i \mathcal{D}c_i e^{-\beta H_s+S_{ret}}
\end{equation}
where $Z_b=Tr e^{-\beta H_b}$ is the partition function for the free photons. $S_{ret}$ takes the form:
\begin{small}
\begin{equation}
S_{ret}=g^2 J^2\int_0^\beta d\tau\int_0^\beta d\tau' \sum_{i,j}J_{i,i+1}(\tau)D(\tau-\tau')J_{j,j+1}(\tau')
\end{equation}
\end{small}
is the action of the photon-induced retarded interaction with the kernel function (in the zero temperature limit):
\begin{equation}
D(\tau-\tau')=\int d \omega \frac{e^{-i\omega(\tau-\tau')}}{\omega^2+\omega_c^2}\sim \frac{1}{\omega_c} e^{-\omega_c(\tau-\tau')}
\end{equation}
In the large $\omega_c$ limit, the retardation effect can be neglected $D(\tau-\tau')\sim \delta(\tau-\tau')$, thus integrating out the photon  leads to an effective Hamiltonian:
 \begin{equation}
 H_{E}=-\frac{(Jg)^2}{\omega_c}\sum_{ij}(c_i^\dag c_{i+1}-c_{i+1}^\dag c_i)(c_j^\dag c_{j+1}-c_{j+1}^\dag c_j), \label{eq:eHam}
 \end{equation}
 which corresponds to an infinite-range correlated hopping for two particles\cite{Chanda2021} {\it e.g.} a pair of fermions simultaneously  hop from the sites $(i+1,j+1)$ to $(i,j)$. This long-range correlated hopping works only when there are more than one fermions in the system.

 {\it Mapping  a 1D 3-body problem to a 3D single-particle problem --} The effective Hamiltonian for $g\ll 1$ become:
 \begin{equation}
 \tilde{H}=H_s+H_{E} \label{eq:Htilde}
 \end{equation}
 where $H_s$ defined in Eq.(\ref{eq:sysHam}) is the unperturbed part, while $H_{E}$ defined in Eq.(\ref{eq:eHam}) is the perturbation. We first diagonalize $H_s$ as $H_s=\sum_i \omega_i \tau_i^\dag \tau_i$, where $\tau_i^\dag$ is the quasiparticle creation operator: $\tau_i^\dag|O\rangle_f=|\bar{i}\rangle$, with $|O\rangle_f$ being the fermion vacuum state, and $|\bar{i}\rangle$ being the $i$th single-particle eigenstate of $H_s$, and $\tau^\dag_i$ can be expressed in terms of a superposition of the fermionic operator $\tau^\dag_i=\sum_i \alpha_i c_i^\dag$  Considering the strong disorder case ($\Delta\gg J$), each single-particle eigenstate of $H_s$ is strongly localized in space, thus $\tau_i^\dag\approx c_i^\dag$ and the effective Hamiltonian.(\ref{eq:Htilde}) turns to:
 \begin{small}
 \begin{equation}
 \tilde{H}\simeq \sum_i \omega_i \tau_i^\dag \tau_i -J'\sum_{ij}(\tau_i^\dag \tau_{i+1}-\tau_{i+1}^\dag \tau_i)(\tau_j^\dag \tau_{j+1}-\tau_{j+1}^\dag \tau_j) \label{eq:Htilde2}
 \end{equation}
 \end{small}
 where $\omega_i\simeq \mu_i$, $J'=\frac{(Jg)^2}{\omega_c}$ and Hamiltonian.(\ref{eq:Htilde2}) is a 1D disordered system with correlated hopping.

 Since the quasi-particle number is also conserved: $\sum_i\tau_i^\dag \tau_i=N_f$, we focus on the subspace of the Hilbert space of Hamiltonian.(\ref{eq:Htilde2}) with $N_f=3$, where the Fock basis can be expressed as $|i,j,k\rangle=\tau_i^\dag \tau_j^\dag\tau_k^\dag|O\rangle_f$, representing a Fock state with the sites $i$, $j$ and $k$ in the 1D lattice being occupied. One can map this three-body Fock basis in 1D lattice to a one-body Fock basis in a 3D lattice: $|i,j,k\rangle\rightarrow |\mathbf{r}\rangle$, where $i$, $j$ and $k$ indicates the $x$, $y$, $z$ coordinate in the 3D lattice respectively. The correlated two-particle hopping discussed above $[i,j,k]\rightarrow[i+1,j+1,k]$ now turns to the single-particle hopping along the face diagonals of the 3D cubic lattice as shown in Fig.\ref{fig:fig1} (b), thus the Hamiltonian.(\ref{eq:Htilde2}) becomes
 \begin{equation}
 \tilde{H}=\sum_\mathbf{r} \Omega_\mathbf{r} C^\dag_\mathbf{r}C_\mathbf{r}-J'\sum_{[\mathbf{r}\mathbf{r}']}[C^\dag_\mathbf{r}C_\mathbf{r'}+h.c] \label{eq:Htilde3}
 \end{equation}
 where $\Omega_\mathbf{r}=\omega_i+\omega_j+\omega_k$, and the summation $[\mathbf{r}\mathbf{r}']$ is over all the face diagonals of the 3D cubic lattice. Hamiltonian.(\ref{eq:Htilde3}) is similar to a 3D Anderson localization model, which exhibits a Mott transition if the disorder strength exceeds a critical value. Such a mapping qualitatively explains the numerical results of the original three-body Hamiltonian.(\ref{eq:ham}). Such a mapping can also tell us what happen in the two-body case $(N_f=2)$, which can be mapped to a 2D disordered system that is always localized, indicating  the cavity mode cannot delocalize the fermions in the two-body case.
 \begin{figure}[htbp]
\centering
\includegraphics[width=0.45\textwidth]{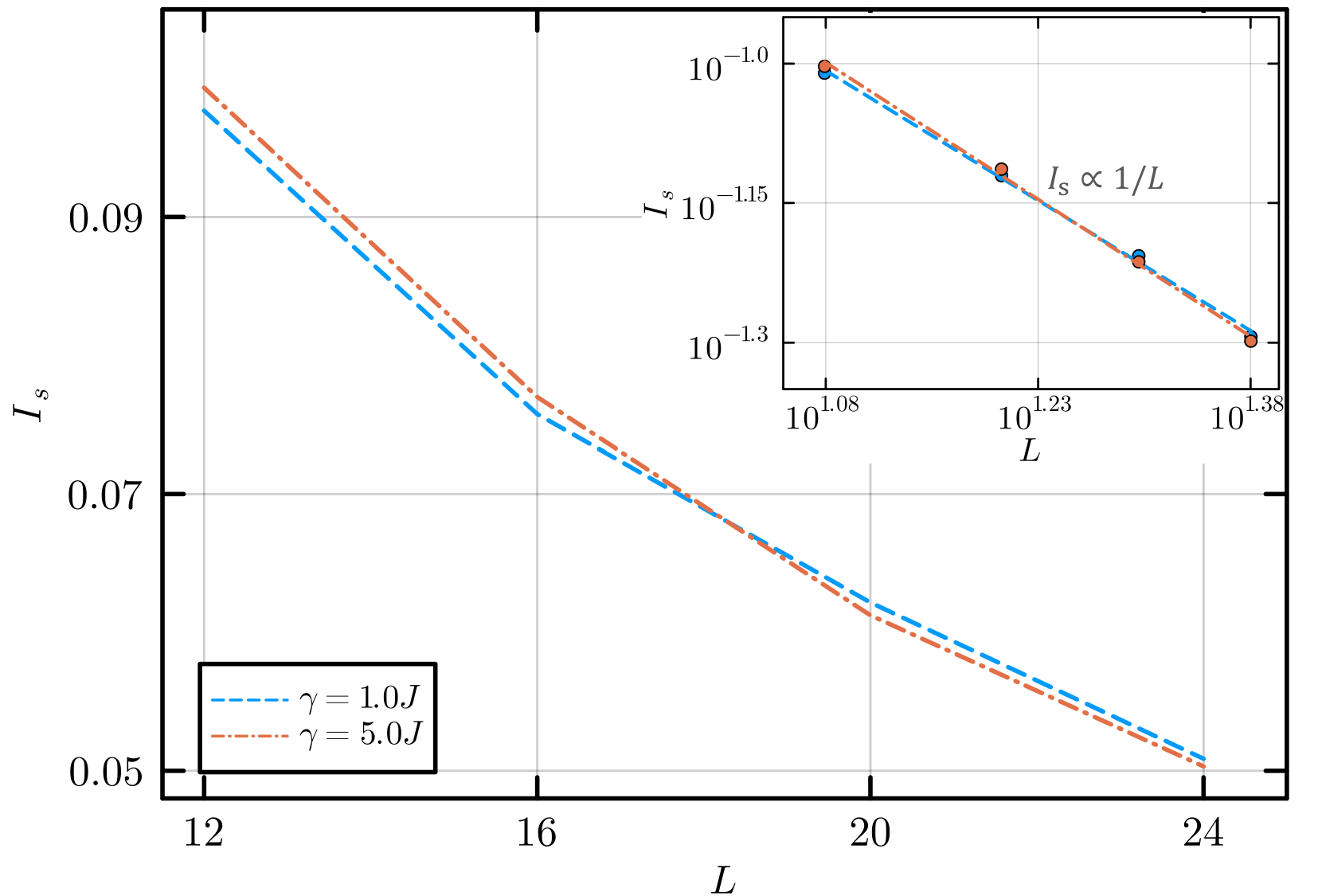}
\caption{The dependence of $I_s$ as defined in Eq.(\ref{eq:Is}) on $L$ in the steady state of the master equation.(\ref{eq:master}) describing the open system with photon leakage. The inset is a log-log plot, which shows that $I_s\sim 1/L$ for different dissipation strength, indicates that the delocalization of fermions persists even in the presence of dissipation. The parameters are chosen as $g=1$, $N_f=3$, $N^b_{max}=2$, $\omega_c=J$, $\Delta=3J$, $\mathcal{N}_{traj}=480$ }
\label{fig:fig4}
\end{figure}

{\it Eigenstate properties and mobility edge --} Next we focus on the eigenstate properties of Hamiltonian.(\ref{eq:ham}) with $N_f=3$. Let $|n\rangle$ be the $n$-th eigenstate of Hamiltonian.(\ref{eq:ham}) with energy $E_n$, to characterize the its localized/extensive feature, we defined a quantity similar to the partition ratio in the case without photon coupling:
\begin{equation}
I_n=\frac1{N_f^2}\sum_i(\langle n|c_i^\dag c_i|n\rangle)^2  \label{eq:In}
\end{equation}
 Under the constraint $\sum_i \langle n|c_i^\dag c_i|n\rangle=3$ $\forall$ n,  $I_n\sim\mathcal{O}(\frac 1L)$   if the particle distribution in $|n\rangle$ is spatially extensive, while $I_n\sim \mathcal{O}(1)$ if it is spatially localized. In Fig.\ref{fig:fig3} (a), we plot $I_n$ as a function of its normalized energy $\epsilon_n=\frac{E_n-E_0}{E_{max}-E_0}$ with $E_0 (E_{max})$ being the lowest (highest) eigen energy of Hamiltonian.(\ref{eq:ham}).  As shown in Fig.\ref{fig:fig3}(b), for those eigenstates in the middle of the spectrum ({\it e.g.} $|n\rangle$ with $\epsilon_n\subset[0.48,0.52]$), the average $I_n$ decay with $L$ as $I_n\sim\frac 1L $, indicating an extensive density distribution. In contrast, for those eigenstates close to the band edge ({\it e.g.} $|n\rangle$ with $\epsilon_n\subset[0,0.04]$), $I_n$ approaches a size-independent constant, a signature of localization.

The localized/delocalized feature of eigenstates can also be captured by the statistics of the ratio of adjacent gaps in different regimes of the energy spectrum\cite{Oganesyan2007}. $r_\alpha=\min(\delta_{\alpha+1},\delta_{\alpha})/\max(\delta_{\alpha+1},\delta_{\alpha})$, with $\delta_\alpha=E_\alpha-E_{\alpha-1}$ are gaps between adjacent energy levels with ordered eigenenergies $\{E_\alpha\}$. The distributions of $r$ in two different regimes of the spectrum are plotted in Fig.\ref{fig:fig3} (c), which exhibit Poisson and GOE distribution respectively, agreeing with localized/delocalized features observed above.  These two distinct types of eigenstates indicate there exists a mobility edge separating the localized and delocalized eigenstates. To verify this point, we plot the average value of $r$ as a function of $\Delta$ and $\epsilon_n$ in Fig.\ref{fig:fig3} (d), which exhibits a boundary between the localized and delocalized eigenstates.

%The quantum entanglement between the fermion and photon is another important quantity to characterize the properties of the eigenstates in such  systems\cite{Passetti2023}.  For a given eigenstate $|n\rangle$, the entanglement entropy is defined as $S_n=-\sum_i \rho_i\ln\rho_i$, where $\rho_i$ are the eigenvalues of the reduced density matrix of the fermion $\hat{\rho}_f=Tr_b |n\rangle\langle n|$ (or boson $\hat{\rho}_b=Tr_f |n\rangle\langle n|$). As shown in Fig.\ref{fig:fig3} (d), for those extensive eigenstates ({\it e.g.} $|n\rangle$ with $\epsilon_n\subset[0.48,0.52]$), the average $S$ approaches a saturated value in the thermodynamic limit. This is due to the fact that the photon couples to the extensive distributed fermions, while the Hilbert space dimension of the photon part takes a finite value ($N_b+1=3$ in our case), which provides an upper bound ($\sim \mathcal{O}(1)$) for the entanglement entropy of its reduced density matrix. On the contrary, for those localized eigenstates ({\it e.g.} $|n\rangle$ with $\epsilon_n\subset[0,0.04]$), the average $S$ decays with system size as $S\sim 1/L$. In this case, photons couple to two spatially localized fermions, but with an effective coupling strength that decays with the system size ($\sim 1/\sqrt{L}$ as shown in Eq.(\ref{eq:ham})). As a consequence, the entanglement between the localized fermions and photons decrease with L.

\begin{figure}[htbp]
\centering
\includegraphics[width=0.45\textwidth]{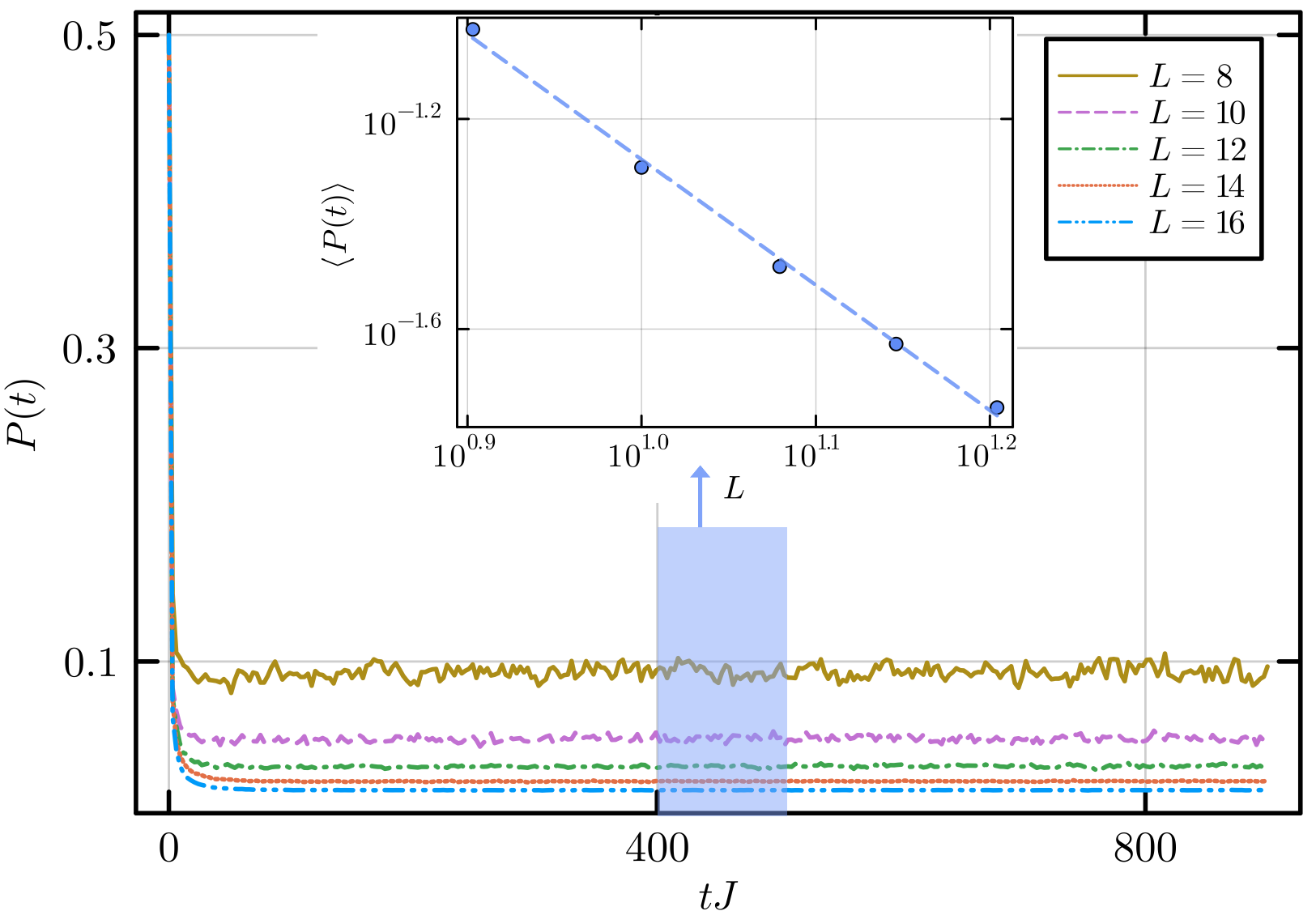}
\caption{Dynamics of population imbalance $P(t)$ in the half-filling  case ($N_f=L/2$) for different system sizes $L$. The inset shows the saturated $P(t)$ averaged over the time period as shown in the blue area as a function of $L$. The parameters are chosen as $g=10$, $N_f=L/2$, $N^b_{max}=2$, $\omega_c=J$, $\Delta=5J$ and $\mathcal{N}_{traj}$ ranges from $10^3$ to $10^6$ depending on $L$.}
\label{fig:fig5}
\end{figure}

{\it Effect of dissipation --} In realistic experimental setups, the leakage of photons from the cavity will result in  dissipation, which can be described by the Lindblad-Markovian master equation:
\begin{equation}
\frac{d\rho}{dt}=i[\rho,H]+\gamma (a\rho a^\dag-\frac 12\{a^\dag a,\rho\}) \label{eq:master}
\end{equation}
where $H$ is the Hamiltonian.(\ref{eq:ham}), $\rho$ is the density matrix of the system (fermion+photon), and $\gamma$ is the dissipation strength. We focus on the steady state of the master equation.(\ref{eq:master}), whose density matrix $\rho_s$ satisfies $\frac{d\rho_s}{dt}=0$. Similar to the closed systems, we define the quantity
 \begin{equation}
 I_s=\sum_i (Tr\rho_s c_i^\dag c_i)^2 \label{eq:Is}
 \end{equation}
 to characterize the spatial  distribution of the fermions in the steady state. As shown in Fig.\ref{fig:fig4}, $I_s$ decay with $L$ as $I_s\sim 1/L$, a signature of extensive distribution,  indicating that in this case, the photon leakage does not qualitatively change the vacuum-induced delocalization.

{\it Many-body generalization --} The correlated hopping proposed above can also significantly change the many-body dynamics of the disordered system. To see this point, we consider the half-filling case for the fermion, and prepare its initial state as a perfect charge-density-wave state as $|\psi\rangle_f=|1_10_2\cdots 1_{L-1}0_L\rangle$, while keeping the photon initial state intact. Starting from such an initial state, we perform the time evolution under the Hamiltonian.(\ref{eq:ham}) and monitor the dynamics of the population imbalance $P(t)=\langle \psi(t)|\frac1L\sum_i(-1)^i c_i^\dag c_i|\psi(t)\rangle$. In a localized phase, the initial state information is kept, thus $P(t)$ will approach a finite value in the long-time limit. In contrast, in the delocalized phase the initial state information will be washed out, thus $P(t)\rightarrow 0$ after sufficient long time. As shown in Fig.\ref{fig:fig5}, for a finite system, $P(t)$ approaches a finite value in the long-time limit, indicating that a proportion of eigenstates are localized. However, as $L$ increases, the saturated value of $P(t)$ decreases algebraically with L (see the inset of Fig.\ref{fig:fig5}), indicating that the proportion of the localized eigenstates decreases with $L$ and all the eigenstates will finally become delocalized in the thermodynamic limit.

{\it Conclusion and outlook --} In summary, a delocalization mechanism based on a correlated hopping is proposed in a disordered cavity quantum system. Throughout this study, we consider the non-interacting fermions, while a natural question is  what happens if the interactions between fermions (e.g. the nearest-neighboring repulsive interactions) are included. Even though the local interactions and cavity-induced long-range interaction/hoppings along are both detrimental to the quantum localization, the interplay between them in a cavity MBL system might lead to richer
phenomena than what is expected on the basis of these effects separately.

{\it Acknowledgments}.---  We thank D. Huse for the stimulating discussions and valuable suggestions about the mapping to a 3D Anderson localization. We also thank Tengzhou Zhang for helpful advices. This work is supported by the National Key Research and Development Program of China (Grant No. 2020YFA0309000), NSFC of  China (Grant No.12174251), Natural Science Foundation of Shanghai (Grant No.22ZR142830),  Shanghai Municipal Science and Technology Major Project (Grant No.2019SHZDZX01).

%\bibliography{vaccum}

\begin{widetext}

%\newpage

\begin{center}
\textbf{\Large{Supplemental material}}
\end{center}

\section{Dynamics of photon}

In the maintext, we mainly focus on the dynamics of the fermions, while here we will study the dynamcis of photon.
As shown in Fig.\ref{npevolute}, in the initial state, system is in the vacuum state of photon, thus the average number of photon $n_p$ starts from zero, and grows during the time evolution. After sufficiently long-time, $n_p$ will saturate to a finite value that depends on the photon-fermion coupling strength. The saturate value is significantly lower than the cut-off number of the photon Hilbert space chosen in the maintext.   We also verify that the increasing of $N_b$ will not qualitatively change the results we discuss in the the main text.

\begin{figure*}[tbhp] \centering
\includegraphics[width=0.6\textwidth]{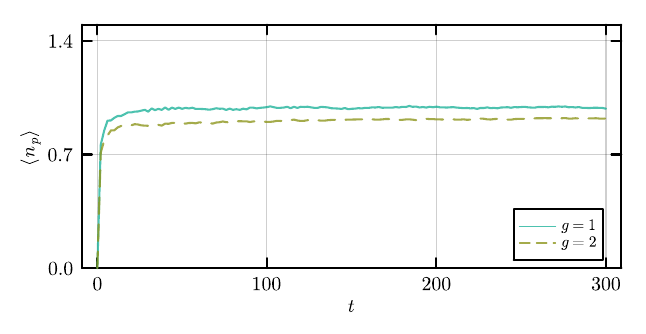}
\caption{(Color online) The time evolution of the photon density under different light-matter coupling strengths when the cutoff $N_b$ is taken as 2. Both cases exhibit quick density saturation considerably lower than the upper bound. Here, while keeping all the other parameters remain unchanged as in Fig.2 (b) in the maintext. We take $L = 100$ and $N_{traj} = 1000$. }\label{npevolute}
\end{figure*}

\section{critical point and critical scaling}

In the main text, we reveal the localized-extensive transition using the partition ratio and energy levels statistics.
However, they rely on obtaining a huge amount of consecutive eigenvalues of the [$C_L^{N} \times  (N_p + 1)$]-dimensional Hamiltonian matrix, and they cannot uncover all the information present in the exact eigenfunctions.
Therefore, we further characterize the transition via a multifractal finite-size scaling analysis\cite{Rudolf2009, Rudolf2010}. The essence of the method is to make use of scale-invariant properties in the vicinity of critical point.

Here, we need to construct a finite-size scaling observable to characterize the multifractality of the critical states.
We first divide the lattice up into regions of $l$ sites, of which there are $(L/l) \equiv \lambda^{-1}$ such regions.
In each region, the coarse-grained weight of the wave function in the region is then given by
\begin{equation}
     \mu_k = \sum_{i = 1}^{l} |\langle i | \psi \rangle|^2 = \sum_{i = 1}^{1/\lambda} \langle \psi | \hat{n}_i |\psi\rangle,
\end{equation}
where $|\psi\rangle$ are taken as the eigenstates with normalized energy close to $\varepsilon$.
The coarse graining enables one to compare systems of different size with fixed $\lambda$.
To be specific, the probability distribution of
\begin{equation}
     \tilde{\alpha}(k) \equiv \log\mu_k/\log\lambda,
\end{equation}
\begin{figure*}[tbhp] \centering
\includegraphics[width=0.5\textwidth]{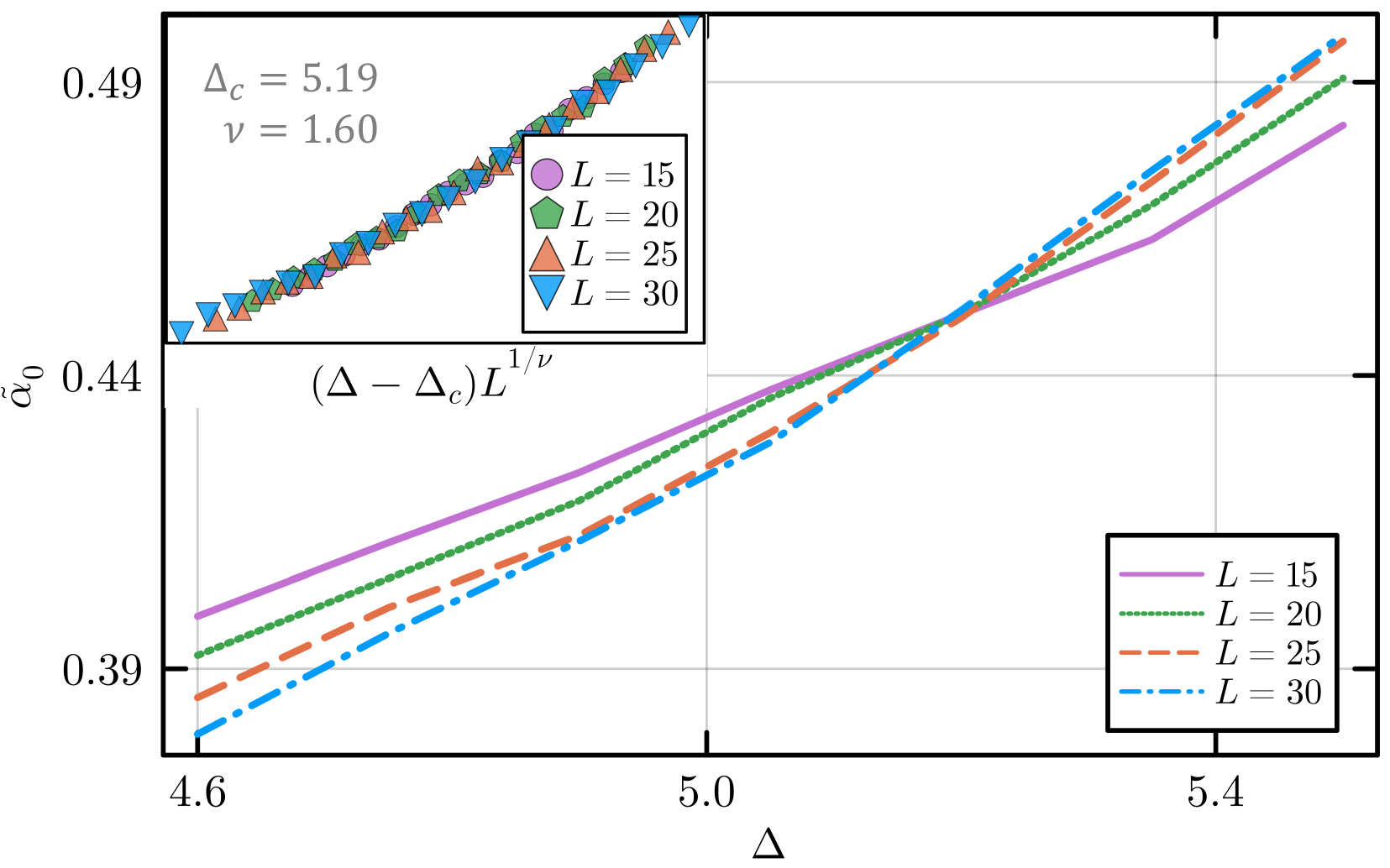}
\caption{(Color online) The quantity $\tilde{\alpha}_0$ versus $\Delta$ for different system sizes near critical point at $\varepsilon = 0.5$. The curves get sharper with increasing system size, and cross at critical point. A standard finite-size scaling estimate for the critical point and exponent gives a scaling exponent estimate $\nu = 1.6$. The scaling collapse is shown in the inset. }\label{scaling}
\end{figure*}
a signature of degree of  localization, is invariant to system size at fixed $\lambda$.
Away from the critical point, this distribution drifts in various directions with $L$ in the localized versus the extensive phase.
Hence,  the mean value of this distribution
\begin{equation}
     \tilde{\alpha}_0 = \langle \log \mu \rangle / \log \lambda
\end{equation}
is an optimal finite-size scaling observable.

Then, we set $\lambda = 0.2$ and $\varepsilon = 0.5$ to investigate the scaling of $\tilde{\alpha}_0$ as a function of $\Delta$ and $L$.
It is notable that we use shift-invert Lanczos to obtain eigenstates near $\varepsilon = 0.5$ with low cost, so that $10^4$ ensemble averages can be taken. Fig.\ref{scaling} shows that the functions $\tilde{\alpha}_0(\Delta)$ get steeper as $L$ increases and there is a crossing at critical point at $\Delta = 5.19$. Assuming the standard simple scaling form $\tilde{\alpha}_0(L, \Delta) = f((\Delta - \Delta_c)L^{1/\nu}$ for some universal function $f$ already allows for an excellent collapse of all the data, as shown in the inset of Fig.\ref{scaling}.
This leads to an estimate for the critical scaling exponent of $\nu \approx 1.6$, which is in good agreement with that of 3D orthongonal Andersion localization transition\cite{Devakul2017}.

\begin{figure*}[tbhp] \centering
\includegraphics[width=0.5\textwidth]{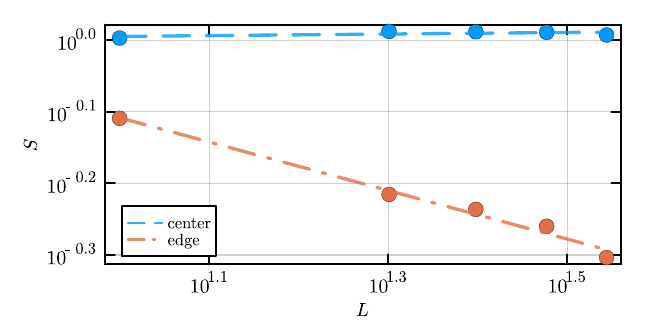}
\caption{(Color online) The average entanglement entropy $S$ between the fermion and photon over the eigenstates in the center($\varepsilon_n \subset$ [0.48, 0.52]) and edge($\varepsilon_n \subset$ [0, 0.04]) regimes as functions of system size $L$. The entanglement entropy of center eigenstates remains unchanged as $L$ approaching thermodynamic limit while that of the edge eigenstates witnesses a power-law decay. Here, we take $N_{traj} = 10^1 - 10^4$, $g = 1$, $\Delta = 4J$ and consider 3-occupied case.}\label{entang}
\end{figure*}

\section{Entanglement entropy analysis}
The quantum entanglement between the fermion and photon is another important quantity to characterize the properties of the eigenstates in such systems\cite{Passetti2023}.
For a given eigenstate $|n\rangle$, the entanglement entropy is defined as
\begin{equation}
    S_n = -\sum_{i} \rho_i \ln \rho_i,
\end{equation}
where $\rho_i$ are eigenvalues of the reduced density matrix of the fermion $\hat{\rho}_f = Tr_b |n\rangle \langle n |$  (or boson $\hat{\rho}_b = Tr_f |n\rangle \langle n |$).
Here, we take $N_{traj} = 10^1 - 10^4$, $g = 1$, $\Delta = 4J$ and consider 3-occupied case. As shown in Fig.\ref{entang}, for those extensive eigenstates (e.g. $\varepsilon_n$ with $\varepsilon_n \subset$ [0.48, 0.52]), the average $S$ approaches a saturated value in the thermodynamic limit. On the contrary, for those localized eigenstates (e.g. $|n\rangle$ with $\varepsilon_n \subset$ [0, 0.04]), the average $S$ exhibits power-law decay with system size.

\end{widetext}
\end{document}